\documentclass{eptcs}
    \newcommand{\IOTS}[1]{\ensuremath{\mathcal{#1}}}
    \newcommand{\ioco}{\mbox{\textbf{ioco}}}
    \usepackage{cancel}
    \newcommand{\notioco}{\ensuremath{\mbox{\textbf{\cancel{ioco}}}}}
    \newcommand{\after}{\mbox{-{after}-}}
    \newtheorem{definition}{Definition}
    \newtheorem{lemma}{Lemma}
    \newtheorem{theorem}{Theorem}
    \newtheorem{corollary}{Corollary}
    \newcommand{\eop}{\ensuremath{\diamond}}
    \newcommand{\tab}{\hspace*{0.5cm}}
    \newcommand{\sinkstate}{\perp}
    \usepackage{graphicx}
    \usepackage[mathcal]{euscript}
    \title{Generating Complete and Finite Test Suite for \ioco: 
Is It Possible?} 
    \newcommand{\titlerunning}{Generating Complete and Finite Test Suite for \ioco: 
Is It Possible?}
    \newcommand{\authorrunning}{Adenilso Simao and Alexandre Petrenko}

\author{
Adenilso Simao
\institute{S\~ao Paulo University}
\institute{S\~ao Carlos, S\~ao Paulo, Brazil}
\email{adenilso@icmc.usp.br}
\and
Alexandre Petrenko
\institute{Centre de recherche informatique de Montreal (CRIM)}
\institute{Montreal, Quebec, Canada}
\email{petrenko@crim.ca}
}

\begin{document}
\maketitle
\begin{sloppypar}
\begin{abstract}Testing from Input/Output Transition Systems has been intensely investigated. The conformance between the implementation and the specification is often determined by the so-called \ioco-relation. However, generating tests for \ioco\ is usually hindered by the problem of conflicts between inputs and outputs. Moreover, the generation is mainly based on nondeterministic methods, which may deliver complete test suites but require an unbounded number of executions. In this paper, we investigate whether it is possible to construct a finite test suite which is complete in a predefined fault domain for the classical \ioco\ relation even in the presence of input/output conflicts. We demonstrate that it is possible under certain assumptions about the specification and implementation, by proposing a method for complete test generation, based on a traditional method developed for FSM. \end{abstract}
    \section{Introduction}
    Testing from Input/Output Transition System (IOTS) has received great attention from academy and industry alike. The main research goal is to devise a theoretically sound testing framework when the behavior of an Implementation Under Test (IUT) is specified as the IOTS model. It is assumed that the tester controls when inputs are applied, while the IUT autonomously controls when, and if, outputs are produced. The IUT's autonomy causes issues in testing. Simply stated, the interaction between the IUT and the tester should be assumed to be asynchronous, since otherwise the tester should have the ability to block the IUT when the latter is ready to produce output but the former has input to be sent. Most approaches based on the so-called \ioco\ conformance relation do not offer sound solutions to the problem of conflicts between inputs and outputs. In particular, the proposal \cite{r15} for input-enabled testers addressing the conflicts lead to uncontrollable tests, while it is widely agreed that only controllable tests, which avoid any choice between inputs or between input and output, should be used. The approaches for test purpose driven test generation from the IOTS implemented in tools such as TGV \cite{r8} and TorX \cite{r16}, as well as in Uppaal Tron which also accepts the IOTS, face the same problem of treating input/output conflicts.
    
    These issues have drawn significant attention of the testing community, e.g., \cite{r1,r6,r7,r11}, and have been dealt with by allowing implicitly or explicitly the presence of channels, e.g., FIFO queues, between the IUT and tester \cite{r6,r7,r17}. However, queues impose a hard burden on the tester, since the communication is now distorted by possible delay in the transmission of messages via queues. In the extreme case, queues render some important testing problems undecidable \cite{r4,r5}. The issue is caused by the conflict between input and output enabled in the same state; while the IUT should be ready to receive input, it may choose to produce an output, blocking or ignoring incoming input. It has been shown that when all the states have either inputs or outputs, but not both, in the so-called Mealy IOTS, such problems do not arise \cite{r12}.\
    
    Apart for the problem of input/output conflicts, the question of generating complete and finite test suite from IOTS w.r.t. the \ioco\ relation remains open. The test generation method which is most referred in the literature relies on non-deterministic choice between: (1) stopping testing; (2) applying a randomly chosen input; or (3) checking for outputs \cite{r14}. The problem with this approach is that, although completeness is guaranteed in some theoretical sense, the practical application of this method is problematic. It requires that the process be repeated an undetermined number of times, since there is no indication of when the completeness has been achieved and thus the process can stop. 
    
    On the other hand, generation methods from Finite State Machines (FSM) approach the problem of test completeness by explicitly stating a set of faulty (mutant) FSMs, called a fault domain, which model potential faults of the IUT; then, a test suite is generated that targets each faulty FSM. Its completeness implies that each IUT possessing the modelled faults will be detected by the test suite. The existing methods for complete test generation are applicable not only to minimal deterministic machines, as the early methods \cite{r2,r3,r18}, but also to nondeterministic FSMs \cite{r10}. This motivated a previous attempt to rephrase FSM methods for checking experiments to the IOTS model \cite{r13}. In particular, an analogue of the Harmonized State Identifier Method (HSI-method) was elaborated there for the trace equivalence relation between the specification and implementation IOTSs. The input/output conflicts were addressed by assuming that the tester detecting (using some means) them will just try to repeatedly re-execute the expected trace to verify if it can be generated by the IUT.
    
    In this paper, we investigate whether it is possible to construct a finite test suite for a given IOTS specification which is complete in a predefined fault domain for the classical \ioco\ relation even in the presence of input/output conflicts. Our solution to the latter is based on the assumption that any IUT in a fault domain resolves each such conflict in favor of inputs; that is, we assume that the IUT is eager to process inputs and, whenever it is in a state where it can either receive an input or produce an output, it will produce an output only if no input is available. We demonstrate this by elaborating a test generation method inspired by the HSI method \cite{r19}, generalizing and adapting its concepts to the realm of IOTS. We illustrate the method with a running example.
    
    This remainder of this paper is organized as follows. In Section 2, we introduce the main concepts of IOTS and test cases. In Section 3, we present the generation method, and demonstrate that the obtained test suite is a complete for a given fault domain. Finally, in Section 4, we conclude the paper and point to future work.
    \section{Input/output transition system and test cases}
    \subsection{Input/output transition system and related definitions}
    
    We use \emph{input/output transition systems} (IOTS, a.k.a. input/output automata \cite{r9}) for modelling systems. Formally, an IOTS \IOTS{S} is a quintuple $(S, s_0, I, O, h_{\IOTS{S}})$, where $S$ is a finite set of states and $s_0 \in S$, is the initial state, $I$ and $O$ are disjoint sets of input and output actions, respectively, and $h_{\IOTS{S}} \subseteq S \times (I \cup O) \times S$ is the transition relation. $\IOTS{S}$ is \emph{deterministic} if $h_{\IOTS{S}}$ is a function on a subset of $S \times (I \cup O)$, i.e., if $(s, x, s') \in h_{\IOTS{S}}$ and $(s, x, s'') \in h_{\IOTS{S}}$, then $s' = s''$. While we shall consider only deterministic IOTSs, they may have output-nondeterminism, i.e., have several outputs enabled in a state. 
    
    For IOTS $\IOTS{S}$, let $init_{\IOTS{S}}(s)$ denote the set of actions enabled at state s, i.e., $init_{\IOTS{S}}(s) = \{x \in I \cup O \mid \exists s' \in S, (s, x, s') \in h_{\IOTS{S}}\}$; let $inp_{\IOTS{S}}(s)$ and $out_{\IOTS{S}}(s)$ denote the set of inputs and outputs, respectively, enabled at state $s$. Thus, $inp_{\IOTS{S}}(s) = init_{\IOTS{S}}(s) \cap I$; $out_{\IOTS{S}}(s) = init_{\IOTS{S}}(s) \cap O$. We omit the subscript if it is clear which IOTS is considered.
    
\begin{figure}
       \centering
       \includegraphics[scale=0.45]{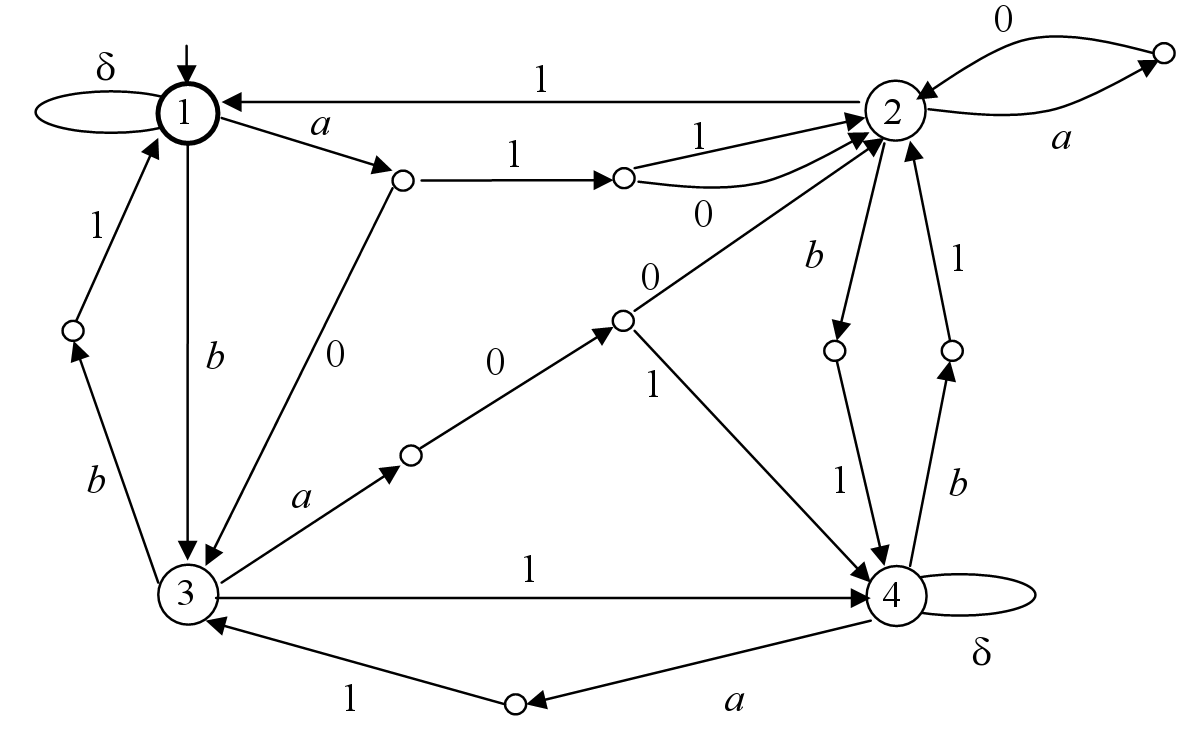}

    \caption{	An IOTS. \label{IOTSM}}
    \end{figure}

    A state $s$ is a \emph{sink} state if $init(s) = \emptyset$; $s$ is an \emph{input state} if $inp(s) \neq \emptyset$. We denote the set of input states by $S_{in}$. An input state s is \emph{stable} (quiescent) if $init(s) \subseteq I$. An input state $s$ is a \emph{quasi-stable} state if $out(s) \neq \emptyset$. In a quasi-stable state, there is an input/output conflict (note that the IOTS itself does not provide any mechanism for resolving such conflicts). A state is an \emph{output state} if it is neither sink nor input state. Figure~\ref{IOTSM} shows an example of an IOTS, where $I = \{a, b\}$ and $O = \{0, 1\}$. Input states are numbered; states $1$ and $4$ are stable, whereas states $2$ and $3$ are quasi-stable.
    
    For IOTS \IOTS{S}, a \emph{path} from state $s_1$ to state $s_{n+1}$ is a sequence of transitions $p = (s_1, a_1, s_2)(s_2, a_2, s_3)\ldots(s_n, a_n, s_{n+1})$, where $(s_i, a_i, s_{i+1}) \in h_{\IOTS{S}}$ for $i = 1, \ldots, n$. Let $\epsilon$ denote the empty sequence of actions. We say that $s_{n+1}$ is \emph{reachable} from $s_1$. IOTS $\IOTS{S}$ is \emph{initially-connected} if each state is reachable from the initial state. A sequence $u \in (I \cup O)^*$ is called a \emph{trace} of $\IOTS{S}$ from state $s_1 \in S$ if there exists path $(s_1, a_1, s_2)(s_2, a_2, s_3)\ldots(s_n, a_n, s_{n+1})$, such that $u = a_1\ldots a_n$. We use the usual operator \emph{after} to denote the state reached after the sequence of actions (we consider only deterministic IOTS), i.e., $s_1{\after}u = s_{n+1}$; if $u$ is not a trace of $s_1$, then $s_1{\after}u = \emptyset$. Let also $Tr(T)$ denote the set of traces from states in $T \subseteq S$. For simplicity, we denote $Tr(\{s\})$ as $Tr(s)$ and use $Tr(\IOTS{S})$ to denote $Tr(s_0)$. A trace $u$ of IOTS $\IOTS{S}$ is \emph{completed}, if $s_0{\after}u$ is a sink state. A trace $u$ of IOTS $\IOTS{S}$ is a \emph{bridge trace from input state} $s$, if $s{\after}u \in S_{in}$ and for each proper prefix $w$ of $u$, $s{\after}w \notin S_{in}$.
    
    Given an IOTS $\IOTS{S} = (S, s_0, I, O, h_{\IOTS{S}})$ and a state $s \in S$, let $\IOTS{S}/s$ denote the IOTS that differs from $\IOTS{S}$ in the initial state changed to $s$, removing states and transitions which are unreachable from $s$.
    
    We use a designated symbol $\delta$ to indicate quiescence in $\IOTS{S}$, that is, the absence of outputs. Quiescence can be encoded by adding self-looping $\delta$ transitions to the stable states; the resulting IOTS has the output action set $O \cup \{\delta\}$. Traces of this IOTS which end with $\delta$ are quiescent traces and traces containing $\delta$ are \emph{suspension} traces. In the rest of the paper, we assume that $Tr(\IOTS{S})$ includes all kinds of traces. 
    
    An IOTS $\IOTS{T} = (T, t_0, I, O, h_{\IOTS{T}})$ is a \emph{submachine} of the IOTS $\IOTS{S} = (S, s_0, I, O, h_{\IOTS{S}})$, if $T \subseteq S$ and $h_{\IOTS{T}} \subseteq h_{\IOTS{S}}$. A state $s \in T$ of a submachine $\IOTS{T}$ of $\IOTS{S}$ is \emph{output-preserving} if for each $x \in O$ such that $(s, x, s') \in h_{\IOTS{S}}$, we have that $(s, x, s') \in h_{\IOTS{T}}$. The submachine $\IOTS{T}$ of $\IOTS{S}$ is \emph{output-preserving} if each state which is not a sink state is output-preserving. The submachine is \emph{trivial} if $T$ is a singleton and $h_{\IOTS{T}} = \emptyset$.
    
    The IOTS $\IOTS{S}$ is \emph{progressive} if it has no sink state and each cycle contains a transition labeled with input, i.e., there is no output divergence. The IOTS $\IOTS{S}$ is \emph{input-complete} if all inputs are enabled in input states, i.e., $inp(s) \neq \emptyset$ implies that $inp(s) = I$, for each state $s$.
    The IOTS $\IOTS{S}$ is \emph{single-input} if $|inp(s)| = 1$, for each input state $s$; it is \emph{output-complete} if $out(s) = O$, for each output state $s$.
    
    In this paper, we assume that specifications and implementations are input-complete progressive deterministic initially-connected IOTS; we let $IOTS(I, O)$ denote the set of such IOTSs with input set $I$ and output set $O$.
     
    To characterize the common behavior of two IOTSs in $IOTS(I, O)$ we use the intersection operation. The \emph{intersection} $\IOTS{S} \cap \IOTS{P}$ of IOTSs $\IOTS{S} = (S, s_0, I, O, h_{\IOTS{S}})$ and $\IOTS{P} = (P, p_0, I, O, h_{\IOTS{P}})$ is an IOTS $(Q, q_0, I, O, h_{\IOTS{S}\cap\IOTS{P}})$ with the state set $Q \subseteq S \times P$, the initial state $q_0 = (s_0, p_0)$, and the transition relation $h_{\IOTS{S}\cap\IOTS{P}}$, such that $Q$ is the smallest state set obtained by using the rule $((s, p), x, (s', p')) \in h_{\IOTS{S}\cap \IOTS{P}} \iff  (s, x, s') \in h_{\IOTS{S}}$ and $(p, x, p') \in h_{\IOTS{P}}$. The intersection $\IOTS{S} \cap \IOTS{P}$ preserves only common traces of both machines; in other words, for each state $(s, p)$ of $\IOTS{S} \cap \IOTS{P}$ we have $Tr((s, p)) = Tr(s) \cap Tr(p)$; moreover, $out((s, p)) = out(s) \cap out(p)$. Thus, $Tr(\IOTS{S} \cap \IOTS{P}) = Tr(\IOTS{S}) \cap Tr(\IOTS{P})$.
     
    Given two IOTSs $\IOTS{S}$ and $\IOTS{T}$, such that $\IOTS{S}$ has at least one sink state $s \in S$, the IOTS obtained by merging the initial state of $\IOTS{T}$ with a sink state s is called the \emph{chaining} of $\IOTS{S}$ and $\IOTS{T}$ in the sink state s, denoted $\IOTS{S} @_s \IOTS{T}$.
    
    For conformance testing, we consider a usual \ioco\ relation.
    \begin{definition}
    Given two IOTSs $\IOTS{P}, \IOTS{S} \in IOTS(I, O)$, $\IOTS{S} = (S, s_0, I, O, h_{\IOTS{S}})$ and $\IOTS{P} = (P, p_0, I, O, h_{\IOTS{P}})$, we write $\IOTS{P}$ \ioco\ $\IOTS{S}$ if for each trace $\alpha \in Tr(\IOTS{S})$, we have that $out(\IOTS{P}{\after}\alpha) \subseteq out(\IOTS{S}{\after}\alpha)$. If $\IOTS{P}$ \ioco\ $\IOTS{S}$ then we say that state $p_0$ is a \emph{reduction} of state $s_0$. The reduction relation between states is also defined for states of the same IOTS $\IOTS{S} \in IOTS(I, O)$, namely, $s_1$ is a \emph{reduction} of $s_2$, if $\IOTS{S}/s_1$ \ioco\ $\IOTS{S}/s_2$.
    \end{definition}
    
    We write $\IOTS{P}$ \notioco\ $\IOTS{S}$, if not $\IOTS{P}$ \ioco\ $\IOTS{S}$. We notice that if the specification IOTS $\IOTS{S}$ contains some state that is a reduction of another state then there exist an implementation $\IOTS{P} \in IOTS(I, O)$ and state $p \in P$, that is a reduction of both states of $\IOTS{S}$. Intuitively, the two states are ``merged'' into a single state in the implementation. As a result, a conforming implementation may have fewer states than its specification. This observation motivates the following definitions and statements.
    
    \begin{definition}
    Two states of $\IOTS{S} \in IOTS(I, O)$ are \emph{compatible}, if there exists a state of an IOTS $\IOTS{P} \in IOTS(I, O)$ that is a reduction of both states; otherwise, i.e., if for any $\IOTS{P} \in IOTS(I, O)$, no state of $\IOTS{P}$ is a reduction of both states, they are \emph{distinguishable}. 
    \end{definition}
    
    According to this definition, compatible states can be ``merged'' in an implementation IOTS into a single state and it can still be a reduction of the specification IOTS, however, any reduction of the specification IOTS cannot have a state that is a reduction of distinguishable states. 
    
    The compatibility of states can be easily determined by the intersection of IOTSs, a simple and inexpensive operation. By definition, if two states of a given IOTS are compatible, there exists a state of some input-complete, progressive IOTS which is a reduction of both states. Such a state is the initial state of the intersection of two instances of a given machine initialized in different states, since the intersection represents all the common traces of the two states. On the other hand, if the two states are distinguishable, the intersection is not a progressive IOTS. This fact is stated in the following lemma.
    
  \begin{lemma}
  Two states $s_1, s_2 \in S$ of $\IOTS{S} = (S, s_0, I, O, h_{\IOTS{S}})$, $\IOTS{S} \in IOTS(I, O)$ are compatible if and only if $\IOTS{S}/s_1 \cap \IOTS{S}/s_2 \in IOTS(I, O)$. 
  \end{lemma}
  
  \textbf{Proof.} Suppose that $s_1$ and $s_2$ are compatible. We show that $\IOTS{S}/s_1 \cap \IOTS{S}/s_2 \in IOTS(I, O)$, that is, $\IOTS{S}/s_1 \cap \IOTS{S}/s_2$ is input-complete, progressive, deterministic and initially-connected. Let $\alpha \in Tr(\IOTS{S}/s_1 \cap \IOTS{S}/s_2)$. Thus, $\alpha \in Tr(\IOTS{S}/s_1) \cap Tr(\IOTS{S}/s_2)$. We have that $s_1' = \IOTS{S}/s_1{\after}\alpha$ and $s_2' = \IOTS{S}/s_2{\after}\alpha$ are also compatible. Hence, by Definition~1, there exists a state $p$ of $\IOTS{P} \in IOTS(I, O)$, with $\IOTS{P} = (P, p_0, I, O, h_{\IOTS{P}})$ that is a reduction of $s_1'$ and $s_2'$. It holds that $out(p) \subseteq out(s_1')$ and $out(p) \subseteq out(s_2')$. As $\IOTS{P}$ is progressive, we have that $init(p) \neq \emptyset$, and thus there exists $x \in out(p)$; hence, $x \in init(s_2') \cap init(s_2')$. It follows that $(\IOTS{S}/s_1 \cap \IOTS{S}/s_2){\after}\alpha$ is not a sink state, since it is followed by $x$, at least. Thus, $\IOTS{S}/s_1 \cap \IOTS{S}/s_2$ has no sink state. If $x$ is an input, then $I \subseteq init(s_1')$ and $I \subseteq init(s_2')$, since $\IOTS{S}$ is input-complete. Therefore, $I \subseteq init((\IOTS{S}/s_1 \cap \IOTS{S}/s_2){\after}\alpha)$, and $\IOTS{S}/s_1 \cap \IOTS{S}/s_2$ is input-complete. As $\IOTS{S}$ is progressive, it does not have cycles with transitions labeled only with outputs. Hence, neither $\IOTS{S}/s_1 \cap \IOTS{S}/s_2$ has such cycles, i.e., $\IOTS{S}/s_1 \cap \IOTS{S}/s_2$ is also progressive. As $\IOTS{S}$ is deterministic and initially-connected, so are $\IOTS{S}/s_1, \IOTS{S}/s_2$ and, consequently, $\IOTS{S}/s_1 \cap \IOTS{S}/s_2$. It follows then that $\IOTS{S}/s_1 \cap \IOTS{S}/s_2 \in IOTS(I, O)$.
  
  Suppose now that the intersection $\IOTS{S}/s_1 \cap \IOTS{S}/s_2 \in IOTS(I, O)$, i.e., it is input-complete, progressive, deterministic and initially-connected. We show that $s_1$ and $s_2$ are compatible, demonstrating that the initial state of $\IOTS{S}/s_1 \cap \IOTS{S}/s_2$ is a reduction of $s_1$ and $s_2$. For each trace $\alpha \in Tr(\IOTS{S}/s_1 \cap \IOTS{S}/s_2)$, we have that $init((\IOTS{S}/s_1 \cap \IOTS{S}/s_2){\after}\alpha) \subseteq init(\IOTS{S}/s_1{\after}\alpha) = init(s_1{\after}\alpha)$; thus, $init((\IOTS{S}/s_1 \cap \IOTS{S}/s_2){\after}\alpha) \cap O = out((\IOTS{S}/s_1 \cap \IOTS{S}/s_2){\after}\alpha) \subseteq init(\IOTS{S}/s_1{\after}\alpha) \cap O = out(\IOTS{S}/s_1{\after}\alpha) = out(s_1{\after}\alpha)$. Therefore, the initial state of $\IOTS{S}/s_1 \cap \IOTS{S}/s_2$ is a reduction of $s_1$. Analogously, $\IOTS{S}/s_1 \cap \IOTS{S}/s_2$ is a reduction of $s_2$ and the result thus follows.\eop
  
  \begin{corollary}
  States $s_1$ and $s_2$ of $\IOTS{S}$ are distinguishable if and only if $\IOTS{S}/s_1 \cap \IOTS{S}/s_2 \notin IOTS(I, O)$, i.e., the IOTS $\IOTS{S}/s_1 \cap \IOTS{S}/s_2$ has a sink state. 
  \end{corollary}
  
  An IOTS in $IOTS(I, O)$ is \emph{input-state-minimal} if every two input states are distinguishable. In the following, we assume that IOTSs which are not input-state-minimal are excluded from $IOTS(I, O)$.
  
  The next lemma states when one state of an IOTS is a reduction of another. The outputs enabled in each state reached in the intersection IOTS, initialized with the respective states, are exactly the outputs enabled in one of the states.
  
  \begin{lemma}
  Given two states $s_1, s_2 \in S$ of $\IOTS{S} = (S, s_0, I, O, h_{\IOTS{S}})$, $s_1$ is a reduction of $s_2$ if and only if $out((s, s')) = out(s)$ for each state $(s, s')$ of $\IOTS{S}/s_1 \cap \IOTS{S}/s_2$.
  \end{lemma}
  \textbf{Proof.} Assume that $s_1$ is a reduction of $s_2$; thus, $\IOTS{S}/s_1$ \ioco\ $\IOTS{S}/s_2$. We have that for each trace $\alpha \in Tr(\IOTS{S}/s_2)$, $out(s_1{\after}\alpha) \subseteq out(s_2{\after}\alpha)$. Let $(s, s')$ be a state of $\IOTS{S}/s_1 \cap \IOTS{S}/s_2$. Thus, there exists a trace $\beta \in Tr(\IOTS{S}/s_1 \cap \IOTS{S}/s_2)$, such that$ (\IOTS{S}/s_1 \cap \IOTS{S}/s_2){\after}\beta = (s, s')$ and, therefore, $\IOTS{S}/s_1{\after}\beta = s$ and $\IOTS{S}/s_2{\after}\beta = s'$. It holds that $\beta \in Tr(\IOTS{S}/s_2)$ and $out(s_1{\after}\beta) \subseteq out(s_2{\after}\beta)$; thus, $out(s) \subseteq out(s')$. We have that $out((s, s')) = out(s) \cap out(s')$. The result then follows, since $out(s) \subseteq out(s')$ and $out((s, s')) = out(s) \cap out(s')$ implies that $out((s, s')) = out(s)$.
  Assume now that $out((s, s')) = out(s)$ for each state $(s, s')$ of $\IOTS{S}/s_1 \cap \IOTS{S}/s_2$. Let $\alpha \in Tr(\IOTS{S}/s_2)$. We have that $\alpha \in Tr(\IOTS{S}/s_1)$ if and only if $\alpha \in Tr(\IOTS{S}/s_1 \cap \IOTS{S}/s_2)$. If $\alpha \notin Tr(\IOTS{S}/s_1)$, then $out(\IOTS{S}/s_1{\after}\alpha) = \emptyset$ and the result follows, since $out(\IOTS{S}/s_1{\after}\alpha) \subseteq out(\IOTS{S}/s_2{\after}\alpha)$. If $\alpha \in Tr(\IOTS{S}/s_1)$, let $(s, s') = \IOTS{S}/s_1{\after}\alpha \cap \IOTS{S}/s_2{\after}a$; thus, $s = \IOTS{S}/s_1{\after}a$ and $s' = \IOTS{S}/s_2{\after}a$. We have that $out(\IOTS{S}/s_1{\after}\alpha \cap \IOTS{S}/s_2{\after}\alpha) = out(\IOTS{S}/s_1{\after}\alpha)$. Let $x \in out(\IOTS{S}/s_1{\after}\alpha)$. As $x \in out(\IOTS{S}/s_1{\after}\alpha \cap \IOTS{S}/s_2{\after}\alpha) = out(\IOTS{S}/s_1{\after}\alpha) \cap out(\IOTS{S}/s_2{\after}\alpha)$, it holds that $x \in out(\IOTS{S}/s_2{\after}\alpha)$. The result then follows, since $out(\IOTS{S}/s_1{\after}\alpha) \subseteq out(\IOTS{S}/s_2{\after}\alpha)$, implying that $\IOTS{S}/s_1$ \ioco\ $\IOTS{S}/s_2$, i.e., $s_1$ is a reduction of $s_2$.\eop
  
    \subsection{Test definitions and problem statement}
  
  To simplify the discussion, we refer to inputs and outputs always taking the view of the implementation, IUT; thus, we say, for instance, that the tester sends an input to the IUT and receives outputs from it, and define test cases accordingly preserving the input and output sets of the specification IOTS $\IOTS{S} = (S, s_0, I, O, h_{\IOTS{S}})$. Recall that $\delta$ is included into $O$; in particular, the output $\delta$ of a test case is interpreted as the fact that the tester executing the test case detects quiescence of the IUT. 
  
  \begin{definition} A test case over input set $I$ and output set $O$ is an acyclic single-input output-complete IOTS $\IOTS{U} = (U, u_0, I, O, h_{\IOTS{U}})$, where $U$ has a designated sink state \emph{fail}. A test case is \emph{controllable} if it has no quasi-stable states, otherwise it is \emph{uncontrollable}. A \emph{test suite} is a finite set of test cases.
  \end{definition}
  
  Let $Tr_{fail}(\IOTS{U})$ be the traces which lead to the sink state $fail$, i.e., $Tr_{fail}(\IOTS{U}) = \{\alpha \in Tr(\IOTS{U}) \mid \IOTS{U}{\after}\alpha = fail\}$. Let $Tr_{pass}(\IOTS{U})$ be the traces which do not lead to $fail$, i.e., $Tr_{pass}(\IOTS{U}) = Tr(\IOTS{U}) \backslash Tr_{fail}(\IOTS{U})$. 
  
  \begin{definition}
  Given the specification IOTS $\IOTS{S}$, a test case $\IOTS{U} = (U, u_0, I, O, h_{\IOTS{U}})$, and an implementation IOTS $\IOTS{B} \in IOTS(I, O)$, 
  \begin{itemize}
\item $\IOTS{B}$ passes the test case $\IOTS{U}$, if the intersection $\IOTS{B} \cap \IOTS{U}$ has no state, where the test $\IOTS{U}$ is in the state fail. 
\item $\IOTS{B}$ fails $\IOTS{U}$, if the intersection $\IOTS{B} \cap \IOTS{U}$ has a state, where the test $\IOTS{U}$ is in the state fail.
\end{itemize}
A test suite $T$ is 
\begin{itemize}
\item \emph{sound for IOTS \IOTS{S} in $IOTS(I, O)$}, if each $\IOTS{B} \in IOTS(I, O)$, such that $\IOTS{B}$ \ioco\ $\IOTS{S}$, passes each test in $T$.
\item \emph{exhaustive for IOTS $\IOTS{S}$ in $IOTS(I, O)$}, if each IOTS $\IOTS{B} \in IOTS(I, O)$, such that $\IOTS{B}$ \notioco\ $\IOTS{S}$, fails some test in $T$.
\item \emph{complete for IOTS \IOTS{S} in $IOTS(I, O)$} w.r.t. the \ioco\ relation, if $T$ is sound and exhaustive for $\IOTS{S}$ in $IOTS(I, O)$.
\end{itemize}
\end{definition}

    Notice that $\IOTS{B}$ passes the test case $\IOTS{U}$, if and only if $Tr(\IOTS{B}) \cap Tr_{fail}(\IOTS{U}) = \emptyset$ and $Tr_{pass}(\IOTS{U}) \subseteq Tr(\IOTS{B})$.
    
    The problem of complete test suite generation for a given IOTS was addressed in \cite{r14,r15}. To generate such a test suite a simple algorithm is suggested, which, however, should be executed an indeterminate number of times to achieve the test completeness w.r.t. the \ioco\ relation. In the first work \cite{r14}, only controllable test cases are generated; the problem with that solution is that the tester must be able to somehow preempt any output each time a test case prescribes sending some input to the IUT. In the second work \cite{r15}, ``the most important technical change with respect to \cite{r14} is the input enabledness of test cases, which was inspired by \cite{r11}''. In terms of our definitions, test cases are uncontrollable; they contain quasi-stable states, where both inputs and outputs are enabled. The intention behind this is to address input/output conflict present in the specification IOTS, since the specification itself provides no clue how an implementation resolves input/output conflict. The behavior of the tester executing uncontrollable test cases may become nondeterministic (the tester has to execute one of the two mutually exclusive actions) and the test results may not always be reproducible. The approaches to generation of controllable tests that tolerate input/output conflicts based on the use of queues are elaborated in several work \cite{r4,r6,r7,r11,r12,r17}. The problem is that one needs to know the size of queues to obtain a finite complete test suite. 
    
    In this paper, we demonstrate, first, that controllable tests that tolerate input/output conflicts can be constructed without knowing the size of queues, and second, that it is possible to obtain in a systematic way a finite set of controllable test cases which is a complete test suite in a finite set of IOTSs. The key assumption we make about the implementation IOTSs in the fault domain is that each implementation when it is a quasi-stable state with the input/output conflict, it does not produce any output if its input queue contains an input. We call such implementations \emph{input-eager}. A subset of $IOTS(I, O)$ that contains input-eager IOTSs is denoted $IEIOTS(I, O)$. Finiteness of complete test suites results from further constraining this set by the number of its input states, as we demonstrate later. 
    
    Testing any input-eager IOTS allows one to use two controllable test cases dealing with input/output conflict; in a quasi-stable state one test case does not send any input and only observes output sequence concluded by quiescence and another one just sends input. In the latter case, the tester does not need to preempt IUT outputs, as an input-eager IOTS will not produce them since the input queue is not empty and contains the input from the tester.
    
    \section{Generating complete test suites for IOTS}
    
  In this section, we investigate whether a classical method for constructing a complete test suite for the FSM model can be reworked to achieve the same result for the IOTS model even with input/output conflicts, namely a test suite with controllable test cases complete in a finite fault domain, without transforming IOTS into Mealy machine. To demonstrate that it is in fact possible, we develop here a counter-part of the HSI-method \cite{r19} for the simplest case, when the FSM is completely specified, minimal, and the fault domain contains FSMs with the number of states not exceeding that of the specification machine.
  
  The HSI-method for FSMs uses sets of distinguishing input sequences, so-called harmonized state identifiers, one per state, such that any two identifiers share an input sequence which distinguishes the two states. These input sequences are appended to state and transition covers in order to check that every state of the implementation corresponds to some state of the specification and every transition of the implementation corresponds to a transition of the specification. 
  
  Accordingly, we need first to define state and transition covers, as well as harmonized state identifiers for a given IOTS.
  
    \subsection{State and transition covers for IOTS}
    
  We first turn our attention to the notion of state cover, needed in tests to eventually establish a mapping from states of the specification to states of the IUT. We focus only on input states of the specification IOTS. First, to check the IUT's reaction to some input it is in fact sufficient to apply the input to a given input state, observe an output sequence, and if it is correct then check whether a proper input state is reached. Output state identification can thus be avoided. However, even considering only input states, some input state of the specification may not be mapped to any state of the IUT even if the latter is a reduction of the specification. Therefore, we should define a state cover targeting only those input states of the specification which have a corresponding state in any \ioco-conforming implementation. 
  
  \begin{definition}
  Given an initially connected IOTS $\IOTS{S}$ and an input state $s$, $s$ is \emph{certainly reachable (c-reachable)}, if any $\IOTS{P} \in IOTS(I, O)$, such that $\IOTS{P}$ \ioco\ $\IOTS{S}$, contains an input state that is a reduction of $s$.
  \end{definition}
  
  It turns out that the certainly reachable states can be determined by considering a submachine of $\IOTS{S}$, similarly to the FSM case \cite{r10}.
  
  \begin{lemma}
  An input state s of an IOTS $\IOTS{S}$ is c-reachable if $\IOTS{S}$ contains a single-input acyclic output-preserving submachine of $\IOTS{S}$ which has $s$ as the only sink state.
  \end{lemma}
  \textbf{Proof.}
  Let $\IOTS{C}_s$ be a single-input acyclic output-preserving submachine of $\IOTS{S}$, which has $s$ as the sink state. The input state $s$ is the only sink state in the submachine; hence all its completed traces converge in $s$. The submachine is output-preserving, this means that for each $\alpha \in Tr(\IOTS{C}_s)$, if $\IOTS{C}_s{\after}\alpha \neq s$ then $out(\IOTS{C}_s{\after}\alpha) = out(\IOTS{S}{\after}\alpha)$. Hence for any IOTS $\IOTS{P} \in IOTS(I, O)$, such that $\IOTS{P}$ \ioco\ $\IOTS{S}$, it also holds that $out(\IOTS{P}{\after}\alpha) \subseteq out(\IOTS{S}{\after}\alpha)$, thus $out(\IOTS{P}{\after}\alpha) \subseteq out(\IOTS{C}_s{\after}\alpha)$. This implies that $\IOTS{P}$ should have at least one of the completed traces of $\IOTS{C}_s$; let $\beta$ be such a completed trace. It is easy to see that $\IOTS{P}$ \ioco\ $\IOTS{S}$ implies that for any $\gamma \in Tr(\IOTS{P})$, $\IOTS{P}{\after}\gamma$ is a reduction of $\IOTS{S}{\after}\gamma$. Hence in any IOTS $\IOTS{P} \in IOTS(I, O)$, such that $\IOTS{P}$ \ioco\ $\IOTS{S}$, the state $\IOTS{P}{\after}\beta$ is a reduction of $\IOTS{S}{\after}\beta$. The result follows, since $\beta$ is a completed trace of $\IOTS{C}_s$ and, thus, $\IOTS{S}{\after}\beta = s$.\eop
  
  \begin{definition}
  Given a c-reachable input state $s$ of an IOTS $\IOTS{S}$, a single-input acyclic output-preserving submachine $\IOTS{C}_s$, which has $s$ as the only sink state, is a \emph{preamble} for state $s$.
  \end{definition}
  
  Preambles for states can be determined by Algorithm 1, adapted from \cite{r10}. 
  
  \noindent\rule{\textwidth}{1pt}
  
  \textbf{Algorithm 1} for constructing a preamble for a given input state. 
  
  \textbf{Input}: An IOTS $\IOTS{S}$ and input state $s \in S$.
  
  \textbf{Output}: a preamble if the state $s$ is c-reachable.
  
  Construct an IOTS $\IOTS{R} = (R, r_0, I, O, h_{\IOTS{R}})$ as follows
  
  $R := \{s\}$;
  
  $h_{\IOTS{R}} := \emptyset$;
  
  While $s_0 \notin R$ and there exist an input state $s' \notin R$ and nonempty $A \subseteq I$, such that for each $x \in A$, $(s', x, s'') \in h_{\IOTS{S}}$, and for each trace $\gamma \in Tr(s'')$, where $\gamma \in O^*$, there exists a prefix $\gamma'$ such that $s''{\after}\gamma' \in R$.
  
  \tab $R := R \cup \{s'\} \cup \{s''{\after}\alpha \mid \gamma \in O*, \gamma \in Tr(s''), \alpha \in pref(\gamma')\}$;
  
\tab $ h_{\IOTS{R}} := h_{\IOTS{R}} \cup \{(s', x, s'') \in h_{\IOTS{S}} \mid x \in A\} \cup \{(s''{\after}\alpha, o, s''{\after}\alpha o) \mid \gamma \in O^*, \gamma \in Tr(s''), \alpha o \in pref(\gamma')\};$

  End While;
  
  If $s_0 \notin R$ then return the message ``the state s is not c-reachable'' and stop;
  
  Else let $\IOTS{R} = (R, r_0, I, O, h_{\IOTS{R}})$, where $r_0 := s_0$, be the obtained IOTS; 
  
  Starting from the initial state, remove in each state all input transitions, but one, to obtain a single-input submachine with the only sink state s;
  
  Delete states which are unreachable from the initial state; 
  
  Return the obtained machine as a preamble for the state $s$ and stop. \eop 
  
  \noindent\rule{\textwidth}{1pt}
  
\begin{figure}
  
  \begin{tabular}{ccc}
  
    \includegraphics[scale=0.45]{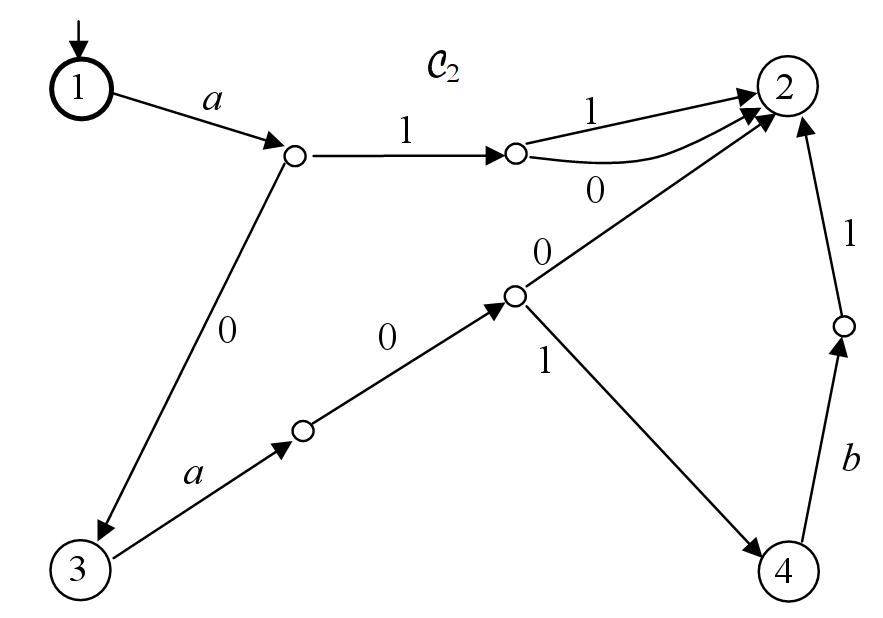} & \mbox{\hspace*{1cm}} &
    \includegraphics[scale=0.45]{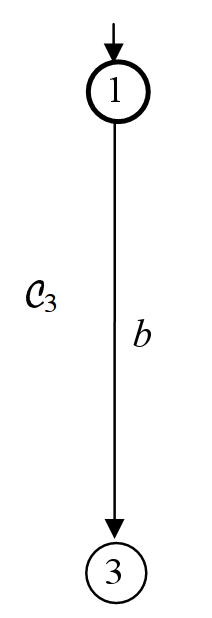}\\

(a) & &							(b)
    \end{tabular}

    \caption{Preambles $\IOTS{C}_2$ and $\IOTS{C}_3$.\label{fig:C2and3}}
    \end{figure}

\begin{figure}
    \centering
    \includegraphics[scale=0.45]{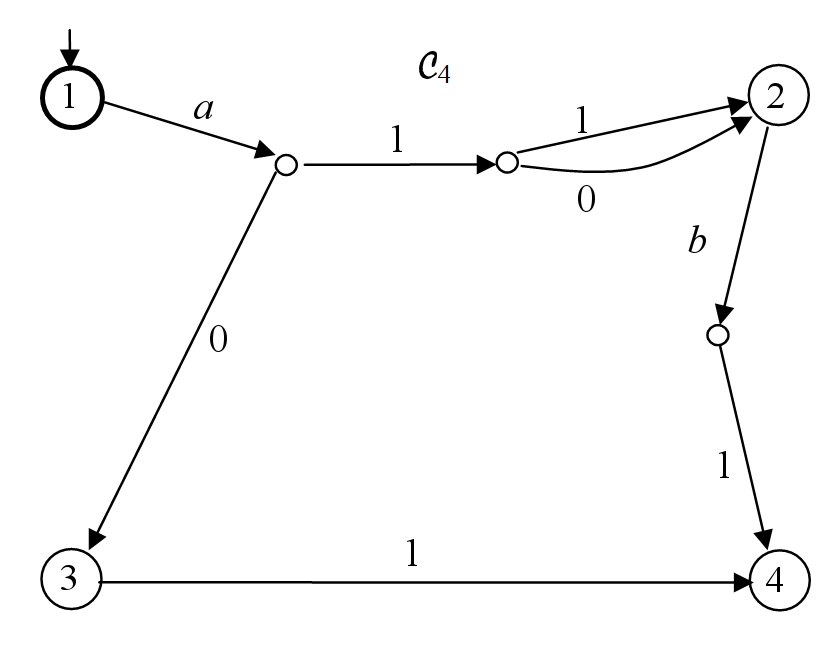}

    \caption{Preamble $\IOTS{C}_4$.\label{fig:C4}}
    \end{figure}

  A preamble can be used to transfer from the initial state to c-reachable input states. For the initial state itself, the preamble is simply the trivial IOTS, which contains only the initial state. Figures~\ref{fig:C2and3}.a, \ref{fig:C2and3}.b and \ref{fig:C4} show the preambles for states 2, 3 and 4, respectively, of the IOTS in Figure ~\ref{IOTSM}.
  
    We assume that each input state of the specification IOTS $\IOTS{S}$ is c-reachable and the initial state is a stable state. An \emph{input state cover} $Z$ of $\IOTS{S}$ is a set of preambles, one for each input state, i.e., $Z = \{\IOTS{C}_s \mid s \in S_{in}\}$.
    
    In FSM-based testing, a state cover is extended to a transition cover, by adding all inputs to each transfer sequence of the state cover. In an IOTS, an input applied in an input state may be followed by a number of output sequences leading to various stable states, creating quiescent traces of IOTS. The set of all possible quiescent traces created by $x \in I$ in input state $s \in S_{in}$ is $\{x\gamma\delta \in Tr(s) \mid \gamma \in O^*\}$. We use $Cov(s, x)$, called \emph{$(s, x)$-cover}, to refer to an IOTS, such that $Tr(Cov(s, x)) = \{x\gamma\delta \in Tr(s) \mid \gamma \in O^*\}$ and the set of sink states is $\{s{\after}x\gamma \mid \gamma \in O^*\}$. For instance, $Cov(2, a)$ for state $2$ and input $a$ of the IOTS in Figure~\ref{IOTSM} has the trace $a01\delta$, whereas $Cov(1, a)$ has the traces $a01\delta$, $a111\delta$ and $a101\delta$.
    A \emph{transition cover} $V$ of $\IOTS{S}$ is the set of preambles of an input state cover chained with $(s, x)$-covers, i.e., $V = \{\IOTS{C}_s @_s Cov(s, x) \mid s \in S_{in}, x \in I\}$.
    Notice that each bridge trace starting from a quasi-stable state $s \in S_{in}$ is covered by $Cov(s', x)$, for some input state $s'$ and input $x$. More generally, we state the following lemma. 
    \begin{lemma}
    Given an IOTS $\IOTS{S} \in IOTS(I, O)$ and a bridge trace $\beta$ from an input state $s \in S_{in}$, there exist input state $s' \in S_{in}$ and input $x$, such that $\gamma\beta\gamma'\delta \in Tr(Cov(s', x))$, for some traces $\gamma \in Tr(s')$ and $\gamma' \in Tr(s'{\after}\gamma\beta)$.
    \end{lemma}
    \textbf{Proof.} If $\beta$ starts with an input, then the results follows directly, since with $\gamma$ as the empty sequence $\beta\gamma'd$ is a quiescent trace starting at state $s$. If $\beta$ starts with an output, then, $\beta \in O^*$ and $s$ is a quasi-stable state. Notice that there exists $\gamma' \in O^*$, such that $\beta\gamma'\delta \in Tr(s)$, since $\IOTS{S}$ is progressive. Moreover, there exist an input state $s'$, a trace $\gamma$ starting with $x$ and followed by outputs, such that $s'{\after}\gamma = s$. Thus, $\gamma\beta\gamma'\delta \in Tr(Cov(s', x))$.\eop
    
    \subsection{State identifiers for IOTS}
    
  The notion of a separator for two states of a given IOTS can be considered as the generalization of the notion of separating sequence used for FSM. 
  
  \begin{definition}
  Given distinguishable states $s_1$ and $s_2$ of an IOTS $\IOTS{S} \in IOTS(I, O)$, a single-input acyclic IOTS $\IOTS{R}(s_1, s_2) = (R, r_0, I, O, h_{\IOTS{R}})$ with the sink states $\sinkstate_{s_1}$ and $\sinkstate_{s_2}$ is a \emph{separator} of states $s_1$ and $s_2$ if the following two conditions hold:
  \begin{itemize}
\item   $r_0{\after}\alpha = \sinkstate_{s_1}$ implies $\alpha \in Tr(s_1) \backslash Tr(s_2)$ and $r_0{\after}\alpha = \sinkstate_{s_2}$ implies $\alpha \in Tr(s_2) \backslash Tr(s_1)$;
\item   for each trace $\alpha$ of $\IOTS{R}(s_1, s_2)$ and input $x$ defined in $r_0{\after}\alpha$, $out(r_0{\after}\alpha x) = out(s_1{\after}\alpha x) \cup out(s_2{\after}\alpha x)$.
  \end{itemize}
  
  The IOTS, obtained by removing from $\IOTS{R}(s_1, s_2)$ the sink state $\sinkstate_{s_2}$ and all transitions leading to it, is called a distinguisher of $s_1$ from $s_2$ and is denoted by $\IOTS{W}(s_1, s_2)$.
  \end{definition}
  
  Separator $\IOTS{R}(s_1, s_2)$ can be obtained from the intersection $\IOTS{S}/s_1 \cap \IOTS{S}/s_2 = (Q, (s_1, s_2), I, O, h_{\IOTS{S}}/s_1\cap {\IOTS{S}}/s_2)$, similar to the case of FSM \cite{r10}, as follows (Algorithm 2). First we determine the intersection $\IOTS{S}/s_1 \cap \IOTS{S}/s_2$ and identify the states where the two IOTSs $\IOTS{S}/s_1$ and $\IOTS{S}/s_2$ disagree on outputs. For each such state, we add transitions leading to sink states $\sinkstate_{s_1}$ and $\sinkstate_{s_2}$. In the final step, we determine a separator as a single-input output-preserving acyclic submachine of the obtained IOTS by removing inputs, as in Algorithm 1.
  
  \noindent\rule{\textwidth}{1pt}
  
  \textbf{Algorithm 2} for constructing a separator for two input states. 
  
  \textbf{Input}: An IOTS $\IOTS{S}$ and distinguishable input states $s_1, s_2 \in S_{in}$.
  
  \textbf{Output}: a separator $\IOTS{R}(s_1, s_2)$.
  
  Construct the IOTS $\IOTS{S}/s_1 \cap \IOTS{S}/s_2 = (Q, (s_1, s_2), I, O, h_{\IOTS{S}/s_1\cap \IOTS{S}/s_2})$
  
  Let $Q_{dis} = \{(s, s') \in Q \mid out(s) \neq out(s')\}$
  
  $h_{dis} = \{((s, s'), o, \sinkstate_{s_1}) \mid (s, s') \in Q_{dis}, o \in out(s) \backslash out(s')\} \cup \{((s, s'), o, \sinkstate_{s_2}) \mid (s, s') \in Q_{dis}, o \in out(s') \backslash out(s)\}$
  
  $h_{\IOTS{P}} = h_{\IOTS{S}/s_1\cap \IOTS{S}/s_2} \cup h_{dis}$
  
  Let $\IOTS{P} = (Q \cup \{\sinkstate_{s_1}, \sinkstate_{s_2}\}, (s_1, s_2), I, O, h_{\IOTS{P}})$
  
  Starting from the initial state, remove in each state all input transitions, but one, to obtain a single-input submachine with the only sink states $\sinkstate_{s_1}$ and $\sinkstate_{s_2}$;
  
  Delete states which are unreachable from the initial state; 
  
  Return the obtained machine as a separator for the states $s_1$ and $s_2$, and stop. \eop
  
  \noindent\rule{\textwidth}{1pt}
  
\begin{figure} 
  \centering
  \begin{tabular}{ccc}
  
    \includegraphics[scale=0.45]{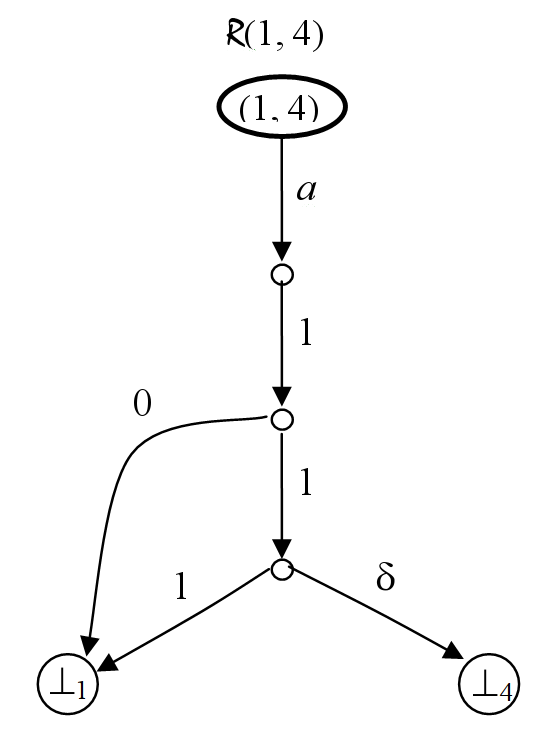} & 
    \includegraphics[scale=0.45]{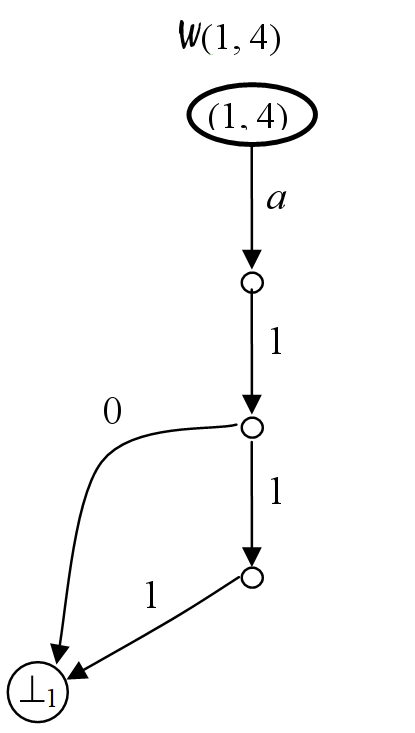} & \includegraphics[scale=0.45]{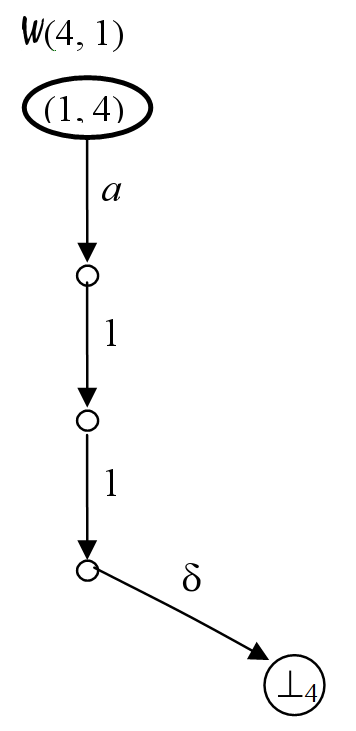}\\

    		(a)		&			(b)		&		(c)
    \end{tabular}
    \caption{(a) Separator $\IOTS{R}(1, 4)$, (b) Distinguisher $\IOTS{W}(1, 4)$ and (c) Distinguisher $\IOTS{W}(4, 1)$.\label{fig:dist}}
    \end{figure}

  Notice that a separator of states $s_1$ and $s_2$ is obviously a separator of $s_2$ and $s_1$, i.e., $\IOTS{R}(s_1, s_2) = \IOTS{R}(s_2, s_1)$, whereas a distinguisher of $s_1$ from $s_2$ is different from a distinguisher of $s_2$ from $s_1$, i.e., $\IOTS{W}(s_1, s_2) \neq \IOTS{W}(s_2, s_1)$. Figure~\ref{fig:dist} shows a separator $\IOTS{R}(1, 4)$ obtained by Algorithm 2, as well as the corresponding distinguishers $\IOTS{W}(1, 4)$ and $\IOTS{W}(4, 1)$.
  
  We consider only input-state-minimal specification IOTS, so we are interested in distinguishers of only input states. If $s_1$ is a stable state and $s_2$ is a quasi-stable state then the separator $\IOTS{R}(s_1, s_2)$ is simple; it has a transition with $\delta$  leading from the state $(s_1, s_2)$ to state $s_1$ and a transition for each $o \in out(s_2)$, leading to $s_2$. Thus, a distinguisher of each stable state from any quasi-stable state has a single $\delta$-transition, we call it a \emph{quiescence distinguisher} of a stable state $s$, denoted $\IOTS{W}^\delta(s)$. It should be included into a stable state identifier of the state $s$.    
  \begin{definition}
  A \emph{state identifier} of input state $s$, denoted $\IOTS{ID}(s)$, is a set of distinguishers $\IOTS{W}(s, s')$ for each input state $s'$ distinguishable from $s$, including $\IOTS{W}^\delta(s)$ if state $s$ is stable.  
  A set of input state identifiers $\{\IOTS{ID}(s) \mid s \in S_{in}\}$, is \emph{harmonized}, if for each pair of input states $s_1$ and $s_2$, such that both are either stable or quasi-stable states, there exists a separator $\IOTS{R}(s_1, s_2)$, such that $\IOTS{W}(s_1, s_2) \in \IOTS{ID}(s_1)$ and $\IOTS{W}(s_2, s_1) \in \IOTS{ID}(s_2)$.
  \end{definition}
  For the IOTS in Figure~\ref{IOTSM}, we have that $\IOTS{ID}(1)$ includes $\IOTS{W}^\delta(1)$ as well as $\IOTS{W}(1, 4)$ in Figure~\ref{fig:dist}. 
    
    \subsection{Complete test suite}
    
\begin{figure}
    \includegraphics[width=0.95\textwidth]{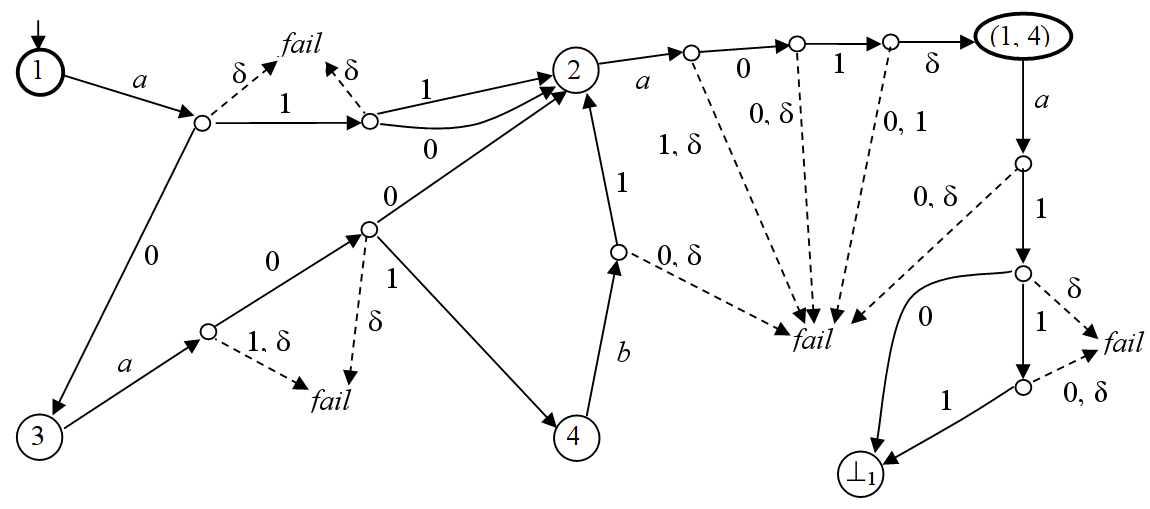}
\caption{Test Case $TC(\IOTS{C}_2 @_2 Cov(2, a) @_1 \IOTS{W}(1, 4))$.\label{fig:tc}}
\end{figure}

    Given the specification IOTS $\IOTS{S} = (S, s_0, I, O, h_{\IOTS{S}})$, $\IOTS{S} \in IOTS(I, O)$, let $Z$ be an input state cover, $V$ be a transition cover of $\IOTS{S}$, and $\{\IOTS{ID}(s) \mid s \in S_{in}\}$ be a set of harmonized identifiers for input states. Consider the set of IOTSs obtained by chaining each IOTS from the input state cover and transition cover with a corresponding harmonized state identifier, namely $D = \{\IOTS{T} @_s \IOTS{R} \mid s \in sink(\IOTS{T}), \IOTS{T} \in (Z \cup V), \IOTS{R} \in \IOTS{ID}(s)\}$, where $sink(\IOTS{T})$ is the set of sink states of $\IOTS{T}$. Each IOTS $\IOTS{U} \in D$ is an acyclic single-input IOTS, since it is obtained by chaining IOTSs with these properties. Moreover, it has no quasi-stable states. If the IOTS $\IOTS{U}$ happens to be also output-complete then it satisfies Definition~3 and is already a test case. The IOTSs in this set can easily be completed with the state $fail$ as follows. Given a single-input acyclic IOTS $\IOTS{U} = (U, u_0, I, O, h_{\IOTS{U}})$, let $TC(\IOTS{U})$ be the IOTS $(T \cup \{fail\}, u_0, I, O, h_{\IOTS{U}} \cup h_f)$, where $h_f = \{(s, o, fail) \mid s \in U, out(s) \neq \emptyset, o \in O \backslash out(s)\}$, which is a test case. Figure~\ref{fig:tc} shows the example of a test case, obtained by chaining the preamble $\IOTS{C}_2$, $Cov(2, a)$ with the quiescent trace $a01\delta$, and distinguisher $\IOTS{W}(1, 4)$. Notice that the quiescence distinguisher $\IOTS{W}^\delta(4)$ of a stable state 4 is also used to identify this state, since the quiescent trace $a01\delta$ has it as a suffix. The $fail$ state is replicated to reduce the clutter.
  
  Completing each IOTS in the set $D$, we finally obtain a test suite $TS = \{TC(\IOTS{U}) \mid \IOTS{U} \in D\}$. Consider now the subset of $IEIOTS(I, O)$ restricted by the number of input states less or equal to that of the specification IOTS $\IOTS{S}$; we denote it by $IEIOTS(I, O, k)$, where $k$ is the number of input states in $\IOTS{S}$. We state the main result of the paper.
  \begin{theorem} 
  Given an IOTS $\IOTS{S} \in IOTS(I, O)$ with $k$ input states, the test suite $TS$ is a complete test suite for $\IOTS{S}$ in $IEIOTS(I, O, k)$ w.r.t. \ioco\ relation.
  \end{theorem}
  
  Before proving Theorem 1, we state some auxiliary results.
  
  \begin{lemma}
  Given two IOTSs $\IOTS{P}$, $\IOTS{S} \in IOTS(I, O)$, if $\IOTS{P}$ is an initially connected submachine of $\IOTS{S}$ with the same initial state $s_0$, then $\IOTS{P}$ \ioco\ $\IOTS{S}$. 
  \end{lemma}
  \textbf{Proof.} Let $\alpha$ be a trace of $\IOTS{S}$. We show that $out(\IOTS{P}{\after}\alpha) \subseteq out(\IOTS{S}{\after}\alpha)$. Let $s = \IOTS{S}{\after}a$. If $s \notin P$, where $P$ is the set of states of $\IOTS{P}$, then $out(\IOTS{P}{\after}\alpha) = \emptyset$, and the result follows. If $s \in P$, we have that $out_{\IOTS{P}}(s) \subseteq out_{\IOTS{S}}(s)$. As $s = \IOTS{P}{\after}\alpha$, the result also follows. Thus, $\IOTS{P}$ \ioco\ $\IOTS{S}$.\eop
  
  \begin{definition}Given two IOTSs $\IOTS{P}$, $\IOTS{S} \in IOTS(I, O)$, $\IOTS{P} = (P, p_0, I, O, h_{\IOTS{P}})$ and $\IOTS{S} = (S, s_0, I, O, h_{\IOTS{S}})$, $\IOTS{P}$ is \emph{input-state homeomorphic} to $\IOTS{S}$, if there exists a bijective map $\varphi$ from $P_{in}$ to $S_{in}$ such that for every state $p \in P_{in}$, each bridge trace $\gamma \in Tr(p)$, it holds that $\varphi(p){\after}\gamma = \varphi(p{\after}\gamma)$.
  
$  \IOTS{P}$ and $\IOTS{S}$ are \emph{input-state isomorphic}, if $\IOTS{P}$ is input-state homeomorphic to $\IOTS{S}$ and $\IOTS{S}$ is input-state homeomorphic to $\IOTS{P}$.
  \end{definition}
  
  Notice that for output-deterministic IOTSs, input-state isomorphic IOTSs are also input-state homeomorphic. An output-nondeterministic IOTS $\IOTS{S}$ that is input-state homeomorphic to $\IOTS{P}$ differs from $\IOTS{P}$ in state names, as well as in the set of bridge traces in some states, since it may have fewer bridge traces, while input-state isomorphic IOTSs differ just in state names.
  
  \begin{corollary} 
  Given two IOTSs $\IOTS{P}, \IOTS{S} \in IOTS(I, O)$, if $\IOTS{P}$ is input-state homeomorphic to $\IOTS{S}$, then $\IOTS{P}$ is input-state isomorphic to an initially connected submachine of $\IOTS{S}$ with $k$ input states and the same initial state.
  \end{corollary}
  \begin{lemma}
  Given an IOTS $\IOTS{S} \in IOTS(I, O)$, let $\IOTS{N} \in IEIOTS(I, O, k)$ be an IEIOTS which passes $TS$. Then $\IOTS{N}$ is input-state homeomorphic to $\IOTS{S}$.
  \end{lemma}
  \textbf{Proof.} Let $\IOTS{N} \in IEIOTS(I, O, k)$, such that $\IOTS{N}$ passes $TS$. $TS$ contains test cases where preambles of an input state cover are chained with harmonized identifiers to the respective states. Thus, for input states $s$ and $s'$, $TS$ contains the test cases $TC(\IOTS{C}_s @_s \IOTS{W}(s, s'))$ and $TC(\IOTS{C}_{s'} @_{s'} \IOTS{W}(s', s))$. Let $\alpha$ be a completed trace of $\IOTS{C}_s$ and $\alpha'$ be a completed trace of $\IOTS{C}_{s'}$, such that $\alpha, \alpha' \in Tr(\IOTS{N})$. As $\IOTS{N}$ passes $TS$, no $fail$ state is reached when the distinguishers $\IOTS{W}(s, s')$ and $\IOTS{W}(s', s)$ are applied after $\alpha$ and $\alpha'$, respectively. Since no state can reach sink state in both distinguishers (see Definition 7), we have that the states $\IOTS{N}{\after}\alpha$ and $\IOTS{N}{\after}\alpha'$ are different, i.e., $\IOTS{N}{\after}\alpha \neq \IOTS{N}{\after}\alpha'$. These are input states, thus, for each pair of input states of $\IOTS{S}$ there exist a pair of distinct states in $\IOTS{N}$; consequently, $\IOTS{N}$ has at least $k$ input states. As $\IOTS{N} \in IEIOTS(I, O, k)$, $\IOTS{N}$ has exactly $k$ input states. 
  
  Let $\IOTS{T} \in (Z \cup V)$, $t \in sink(\IOTS{T})$, $\alpha \in Tr(\IOTS{T})  \cap Tr(\IOTS{N})$, such that $\IOTS{T}{\after}\alpha = t$, $\IOTS{N}{\after}\alpha \in N_{in}$. Similarly, let $\IOTS{T}' \in (Z \cup V)$, $t' \in sink(\IOTS{T}')$, $\alpha' \in Tr(\IOTS{T}')  \cap Tr(\IOTS{N})$, such that $\IOTS{T}'{\after}\alpha' = t'$. Notice that $\alpha$ and $\alpha'$ are completed traces of IOTSs in the state or transition cover, which are also traces of $\IOTS{N}$. We prove that $\IOTS{S}{\after}\alpha' = \IOTS{S}{\after}\alpha$  if and only if $\IOTS{N}{\after}\alpha' = \IOTS{N}{\after}\alpha$. Let $s = \IOTS{S}{\after}\alpha$ and $s' = \IOTS{S}{\after}\alpha'$. Suppose first that $\IOTS{S}{\after}\alpha' \neq \IOTS{S}{\after}\alpha$. Thus, TS contains $TC(\IOTS{T} @_s \IOTS{W}(s, s'))$ and $TC(\IOTS{C}_{s'} @_{s'} \IOTS{W}(s', s))$, and as $\IOTS{N}$ passes $TS$, no $fail$ state is reached when the distinguishers $\IOTS{W}(s, s')$ and $\IOTS{W}(s', s)$ are applied after $\alpha$ and $\alpha'$, respectively. Since no state can reach sink state in both distinguishers, we have that $\IOTS{N}{\after}\alpha \neq \IOTS{N}{\after}\alpha'$. Suppose now that $\IOTS{S}{\after}\alpha' = \IOTS{S}{\after}\alpha$. We prove by contradiction that $\IOTS{N}{\after}\alpha' = \IOTS{N}{\after}\alpha$. Assume that $\IOTS{N}{\after}\alpha' \neq \IOTS{N}{\after}\alpha$. Thus, let $s''$ be an input state, different from $s = \IOTS{S}{\after}\alpha$. Let $\beta \in Tr(\IOTS{C}_{s''})$, such that $\beta \in Tr(\IOTS{N})$. As $TS$ contains $TC(\IOTS{T} @_s \IOTS{W}(s, s''))$ and $TC(\IOTS{C}_{s''} @_{s''} \IOTS{W}(s'', s))$ and $\IOTS{N}$ passes $TS$, we have that $\IOTS{N}{\after}\alpha \neq \IOTS{N}{\after}\beta$. Analogously, we can show that we have that $\IOTS{N}{\after}\alpha' \neq \IOTS{N}{\after}\beta$. Thus, $\IOTS{N}{\after}\alpha$ is distinct from $k - 1$ distinct input states of $\IOTS{N}$ and $\IOTS{N}{\after}\alpha'$ is also distinct from $k - 1$ distinct input states of $\IOTS{N}$. As $\IOTS{N}{\after}\alpha' \neq \IOTS{N}{\after}\alpha$, $\IOTS{N}$ has $k + 1$ states, which contradicts the fact that $\IOTS{N} \in IEIOTS(I, O, k)$ and has at most $k$ input states. Therefore, $\IOTS{N}{\after}\alpha' = \IOTS{N}{\after}\alpha$. 
  Thus, let $\varphi$ be a bijection from the input states $N_{in}$ of $\IOTS{N}$ to the input states $S_{in}$ of $\IOTS{S}$, such that for each completed trace $\chi$ of an IOTS in the state cover $Z$ or transition cover $V$, which is also a trace of $\IOTS{N}$, we have that $\varphi(\IOTS{N}{\after}\chi) = \IOTS{S}{\after}\chi$. 
  Let $p$ be an input state of $\IOTS{N}$. There exists a completed trace $\alpha$ of an IOTS in the input state cover $Z$, such that $\alpha$ is also a trace of $\IOTS{N}$ and $\IOTS{N}{\after}\alpha = p$. Thus, it holds that $\varphi(\IOTS{N}{\after}\alpha) = \varphi(p) = \IOTS{S}{\after}\alpha$. Let $\gamma \in Tr(p)$ be a bridge trace, such that $\alpha\gamma$ is a completed trace of an IOTS in the transition cover $V$. Thus, it follows that $\varphi(p){\after}\gamma = \varphi(\IOTS{N}{\after}\alpha){\after}\gamma = (\IOTS{S}{\after}\alpha){\after}\gamma = \IOTS{S}{\after}\alpha\gamma = \varphi(\IOTS{N}{\after}\alpha\gamma) = \varphi((\IOTS{N}{\after}\alpha){\after}\gamma) = \varphi(p{\after}\gamma)$, i.e., $\varphi(p){\after}\gamma = \varphi(p{\after}\gamma)$. Therefore, we have that $\IOTS{N}$ is input-state homeomorphic to $\IOTS{S}$.\eop
  
  We can now prove Theorem 1.
  
  \textbf{Proof of Theorem 1}. We first prove that $TS$ is sound for $\IOTS{S}$ in $IEIOTS(I, O, k)$. Let $\IOTS{N} \in IEIOTS(I, O, k)$, such that $\IOTS{N}$ \ioco\ $\IOTS{S}$. We have that for each test $\IOTS{U} \in TS$, $Tr_{pass}(\IOTS{U}) \subseteq Tr(\IOTS{S})$. Thus, $Tr_{pass}(\IOTS{U} \cap \IOTS{S}) = Tr_{pass}(\IOTS{U}) \cap Tr(\IOTS{S}) = Tr_{pass}(\IOTS{U})$. Since $\IOTS{N}$ \ioco\ $\IOTS{S}$, we have, for each $\alpha \in Tr(\IOTS{S})$, $out(\IOTS{N}{\after}\alpha) \subseteq out(\IOTS{S}{\after}\alpha)$. Let $\beta \in Tr_{pass}(\IOTS{U} \cap \IOTS{N})$; hence, $\beta \in Tr_{pass}(\IOTS{U})$ and $\beta \in Tr(\IOTS{N})$. As $Tr_{pass}(\IOTS{U}) \subseteq Tr(\IOTS{S})$, we have that $\beta \in Tr(\IOTS{S})$. It follows that $Tr_{pass}(\IOTS{U} \cap \IOTS{N}) = Tr_{pass}(\IOTS{U}) \cap Tr(\IOTS{N}) \subseteq Tr_{pass}(\IOTS{U}) \cap Tr(\IOTS{S}) = Tr_{pass}(\IOTS{U} \cap \IOTS{S}) = Tr_{pass}(\IOTS{U})$. Hence, $Tr_{pass}(\IOTS{U} \cap \IOTS{N}) \subseteq Tr_{pass}(\IOTS{U})$. As a result, $\IOTS{N}$ passes each test of $TS$, and $TS$ is thus sound for $\IOTS{S}$ in $IEIOTS(I, O, k)$ for the \ioco\ relation.
  
  We now prove by contradiction that $TS$ is exhaustive for $\IOTS{S}$ in $IEIOTS(I, O, k)$. Assume that $TS$ is not exhaustive $\IOTS{S}$ in $IEIOTS(I, O, k)$; thus, there exists $\IOTS{N} \in IEIOTS(I, O, k)$, such that $\IOTS{N}$ \notioco\ $\IOTS{S}$ and $\IOTS{N}$ passes $TS$. As $\IOTS{N}$ passes $TS$, by Lemma 6, we have that $\IOTS{N}$ is input-state homeomorphic to $\IOTS{S}$; thus, by Corollary 2, $\IOTS{N}$ is input-state isomorphic to an initially connected submachine of $\IOTS{S}$ with $k$ input states; hence, by Lemma 5, $\IOTS{N}$ \ioco\ $\IOTS{S}$, a contradiction. We conclude then that $TS$ is exhaustive for $\IOTS{S}$ in $IEIOTS(I, O, k)$.
  
  Therefore, $TS$ is complete for $\IOTS{S}$ in $IEIOTS(I, O, k)$ w.r.t. the \ioco\ relation.\eop
  
    \section{Concluding Remarks}
    
  In this paper, we have investigated whether it is possible to construct a finite test suite for a given IOTS specification which is complete in a predefined fault domain for the classical \ioco\ relation even in the presence of input/output conflicts. Our conclusion is that it is in fact possible; however, under a number of assumptions about the implementations and the specifications. We have proposed a generation method which produces a finite test suite, which is complete for a given fault domain. The issue of conflicts between inputs and outputs is tackled by assuming that the implementation is ``eager'' to read inputs and thus such conflict is solved in favor of input, i.e., outputs are produced only if no input is presented to the implementation.
  
  The proposed generation method is based on a classical FSM method. Thus, we rephrased the notions related to FSM generation methods, such as state cover, transition cover, state identifier, to the IOTS model. The method applies to IOTS that is minimal in the sense defined in the paper and each input state is reachable in any \ioco-conforming implementation. A remarkable feature of the method is that it requires no assumption about distinguishability of output states or about their number in the specification and any implementation. Also no bound on the buffer's length in the implementation is required to generate a complete test suite.  
  
  Our future work will focus on extending the class of IOTSs for which the approach is applicable by relaxing the mentioned constraints.
  
\section*{Acknowledgment} The first author would like to thank Brazilian Funding Agency FAPESP for its partial financial support (Grant 12/02232-3). We would like to thank the anonymous reviewers for the suggestions that helped improving the paper.

\bibliography{test-gen-iots}
\end{sloppypar}
\end{document}